\documentclass[b5paper,twoside]{jpconf}  
\usepackage{graphicx}

\begin{document}

\title[Single Star Scidar]{The Development of Single Star Scidar for Tibet and Dome A}

\author[L.Y. Liu et al.]{Li-Yong Liu$^1$$^,$$^2$, Yong-Qiang Yao$^1$$^,$$^2$, Jean Vernin$^3$, Hong-Shuai Wang$^1$$^,$$^2$, Jia Yin$^1$$^,$$^2$, Xuan Qian$^1$$^,$$^2$}

\address{$^1$National Astronomical Observatories, Chinese Academy of Sciences, Beijing 100012, China}
\address{$^2$Key Laboratory of Optical Astronomy, Chinese Academy of Sciences, Beijing 100012, China}
\address{$^3$Laboratoire H. Fizeau, UMR 6525, Universit\'e de Nice-Sophia Antipolis, Observatoire de la C\^ote d'Azur, Parc Valrose, 06108 Nice Cedex 2, France}

\ead{liuly@nao.cas.cn}

\begin{abstract}

A Single Star Scidar system(SSS) has been developed for remotely
sensing atmospheric turbulence profiles. The SSS consists of
computing the spatial auto/cross-correlation functions of short
exposure images of the scintillation patterns produced by a
single star, and provides the vertical profiles of optical
turbulence intensity $C_n^2(h)$ and wind speed $V(h)$.
The SSS needs only a 40$cm$ aperture telescope,
so that can be portable and equipped easily to field candidate sites.
Some experiments for the SSS have been made in Beijing last year,
successfully retrieving atmospheric turbulence and wind profiles from the ground to 30$km$.
The SSS observations has recently been made at the Xinglong station of NAOC,
characterizing atmospheric parameters at this station.
We plan to automatize SSS instrument and run remote observation via internet;
a more friendly auto-SSS system will be set up and make use at the candidate sites in Tibet and Dome A.

\end{abstract}

\section{Introduction}

The atmospheric transparency and stability are the preferred
indicators for the quality of ground-based astronomical observatory.
Atmospheric optical turbulence seriously affects the performance of
large optical-infrared telescopes. The development of detection
technique to atmospheric optical turbulence will provide with
scientific judgment for site selection and evaluation, and also with
valuable database for other research projects, such as design of
adaptive optical system, atmospheric optical communication, and
detection of space objects.

Because the profiles of atmospheric optical turbulence is very important
to evaluate astronomical sites, we have performed investigation into
the related techniques, and finally made decision to develop an
advanced instrument called Scidar (SCIntillation Detection and
Ranging)\cite{vernin83a,vernin83b}. Scidar is one technology at most
in details for monitoring optical turbulence, and has been employed on some existing observatories.
Scidar can detect multiple layers of atmospheric turbulence, with much higher spatial resolution,
which is very important for site evaluation and adaptive optics application.
Previously, the disadvantage of Scidar technique is aperture requirement for a telescope, usually larger than 1$m$,
and it is difficult to use on field candidate sites in the early stage of site testing.

Our aim is to develop a Single Star Scidar\cite{habib06},
the most advanced technology currently for measuring vertical profile of optical turbulence.
This SSS needs only a 40$cm$ telescope, which make it  portable and equipped easily
to the field candidate sites, such as in west China and Antarctic Dome A.
The monitoring results will provide
knowledge of detailed turbulence profiles on-site for the next generation large telescopes of China.

\section{Instrument}

Fig.1 shows the SSS sysytem installed at the Xinglong station of
National Astronomical Observatories, CAS. The SSS configuration is
as the following:
1) a 40$cm$ Meade LX200 tube on an Astro-Physics 1200 equatorial mount;
2) a collimating lens at the focus of the telescope, to make the beam parallel;
3) a 640$\times$480 Pixelfly CCD camera, attached under the collimating lens for fast sampling star scintillation pattern;
4) an auto guide system with a LPI CCD.

\begin{figure}[h]
\begin{minipage}{15pc}\hspace{1pc}
\includegraphics[width=10pc,angle=270]{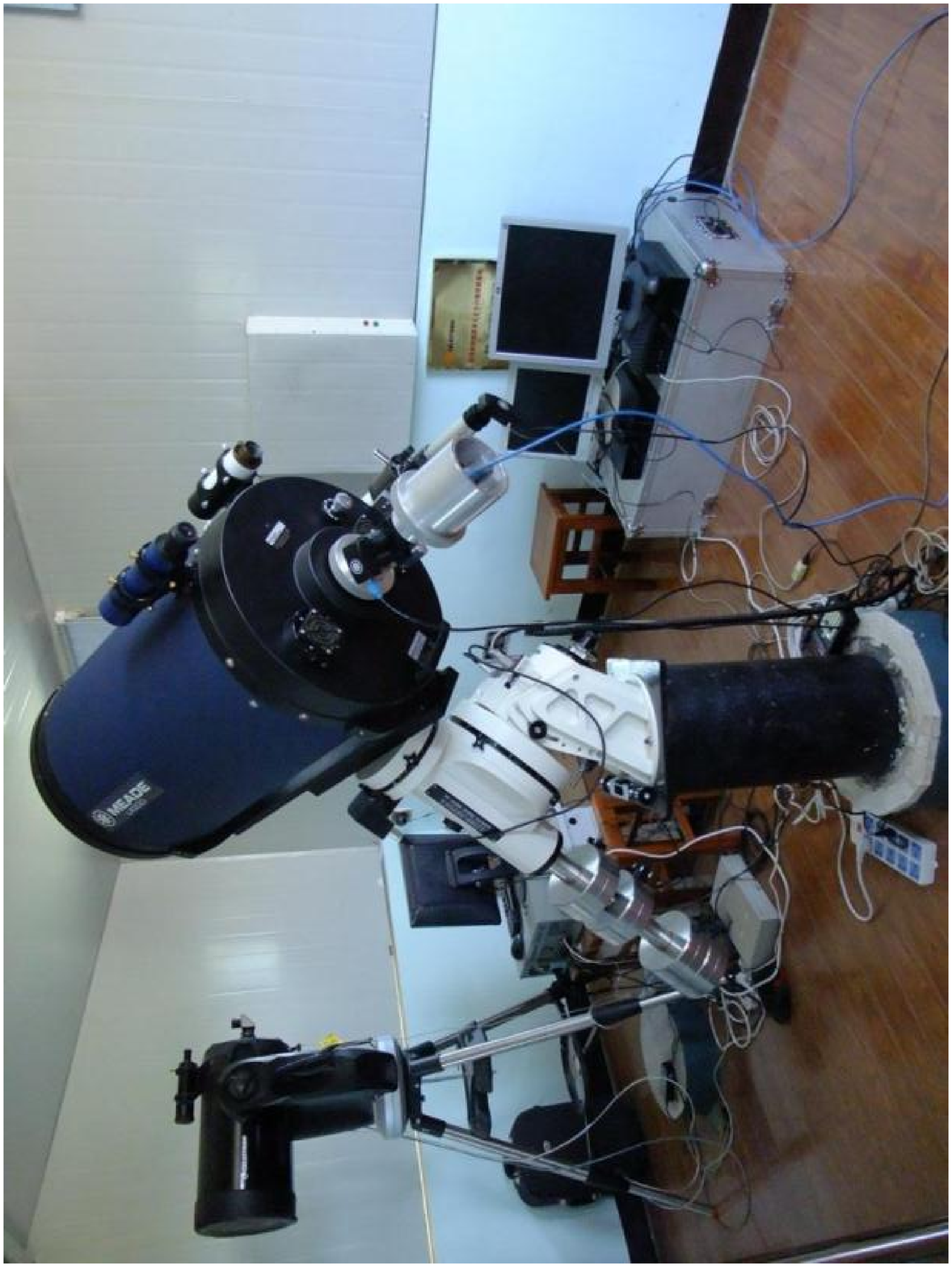}
\caption{\label{label}The SSS system installed at Xinglong Station.}
\end{minipage}\hspace{1.5pc}%
\begin{minipage}{15pc}
\includegraphics[width=10pc,angle=270]{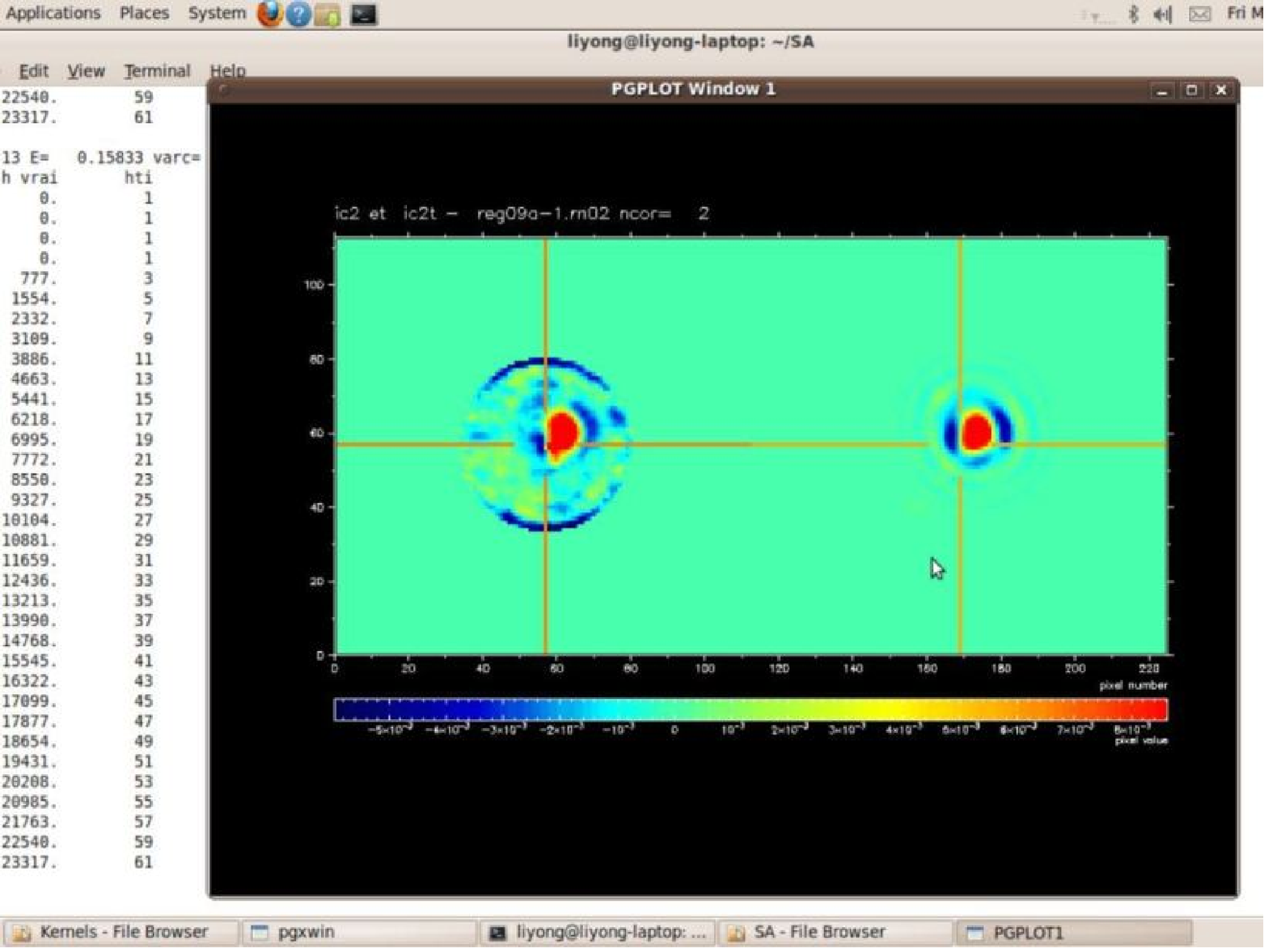}
\caption{\label{label}The ``simulated annealing'' process for reconstructing
observation results.}
\end{minipage}
\end{figure}

The SSS is a new technique for retrieving atmospheric turbulence
profile by analysis of single star scintillation.
The scintillation patterns are analyzed in real time, by computing the spatial
auto-correlation and at least two cross-correlation images performed
by a dual core computer. Fig.2 displays the results of temporal
cross-correlation of scintillation pattern at 2$\Delta$$T$(left),
and the same function reconstructed with simulated annealing algorithm(right).
The off-line processes let assess both optical turbulence and wind speed profiles
using the simulated annealing method.

\section{Observations and Results}

The SSS instrument was constructed and at first made experiments in
Beijing, and then shipped to Xinglong Station, which is located at
N40$^\circ$23$\verb|'|$, E117$^\circ$35$\verb|'|$, 950$m$ high and
170 miles northeast to Beijing. The SSS observations were carried
out in several periods from April to September, 2011. We have
successfully obtained the profiles of optical turbulence and wind
speed, as well as the key parameters of optical turbulence, seeing,
coherence length, coherence time, and isoplanatic angle.

\begin{figure}[h]
\begin{minipage}{14pc}
\includegraphics[width=11.5pc,angle=270]{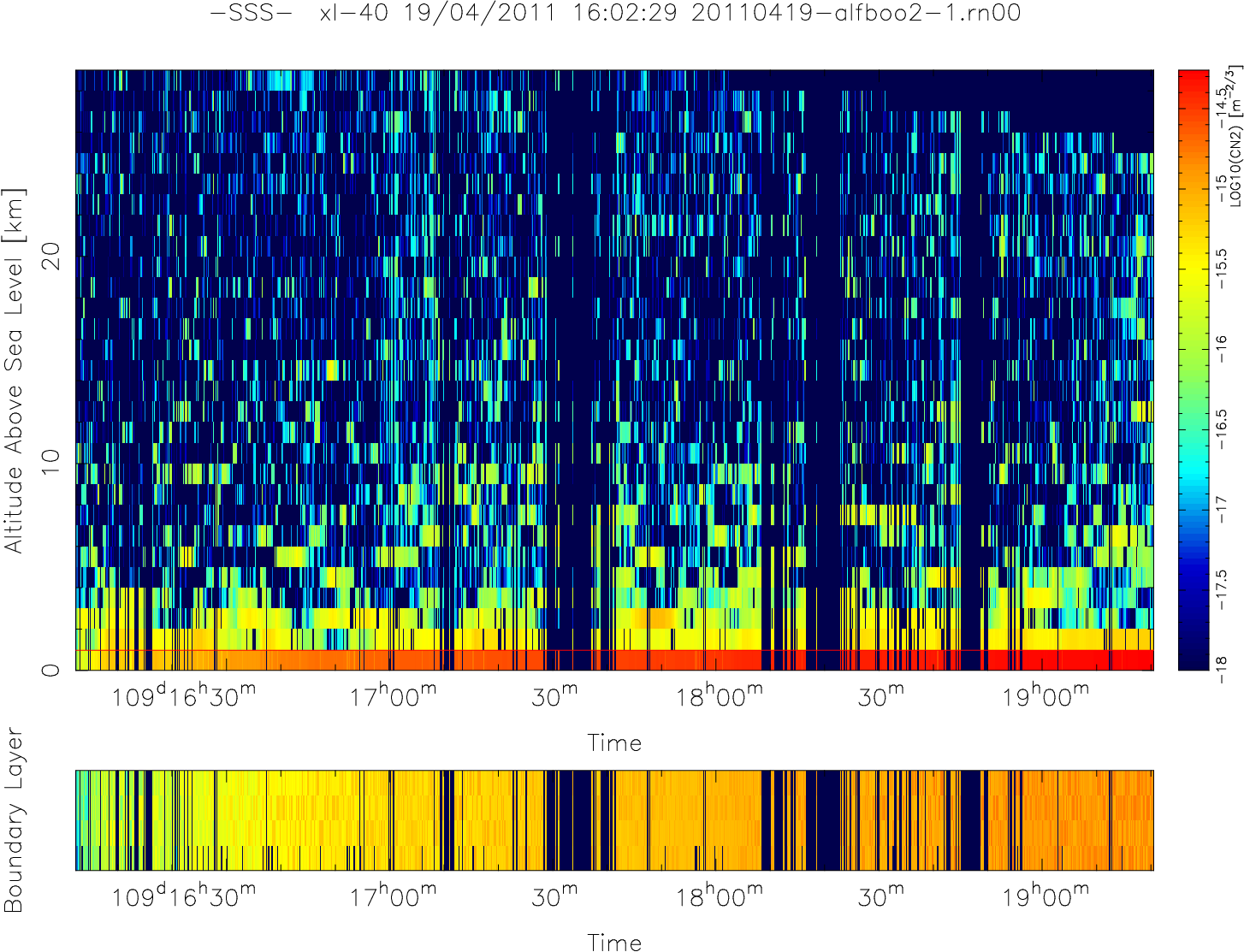}
\caption{\label{label}Temporal evolution of $C_n^2(h,t)$ on 19 April
2011.}
\end{minipage}\hspace{2.5pc}%
\begin{minipage}{14pc}
\includegraphics[width=11.5pc,angle=270]{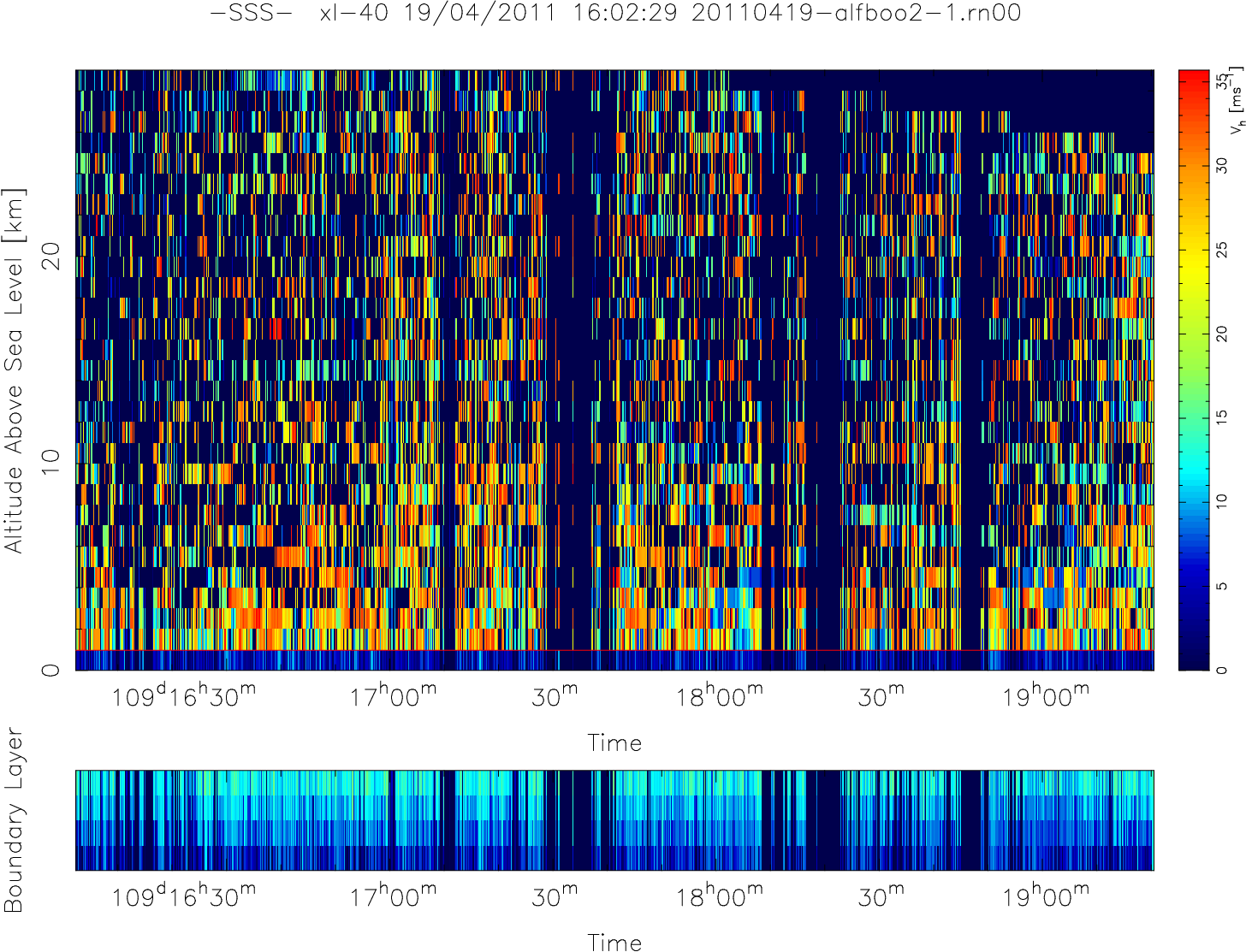}
\caption{\label{label}Temporal evolution of $|V(h,t)|$ on 19 April
2011.}
\end{minipage}
\end{figure}

Fig.3 and Fig.4 are the temporal evolution of the optical turbulence
and the wind speed modulus simultaneously. Fig.5 shows the median
vertical profiles of turbulence($top-left$), the frequency counts of
detected layers($top-right$); the profile of mean speed modulus and
mean wind direction($bottom-left$), and the mean vertical profile of
the wind speed standard deviation($bottom-right$).

\begin{figure}
\begin{center}
\includegraphics[width=23pc,angle=-90]{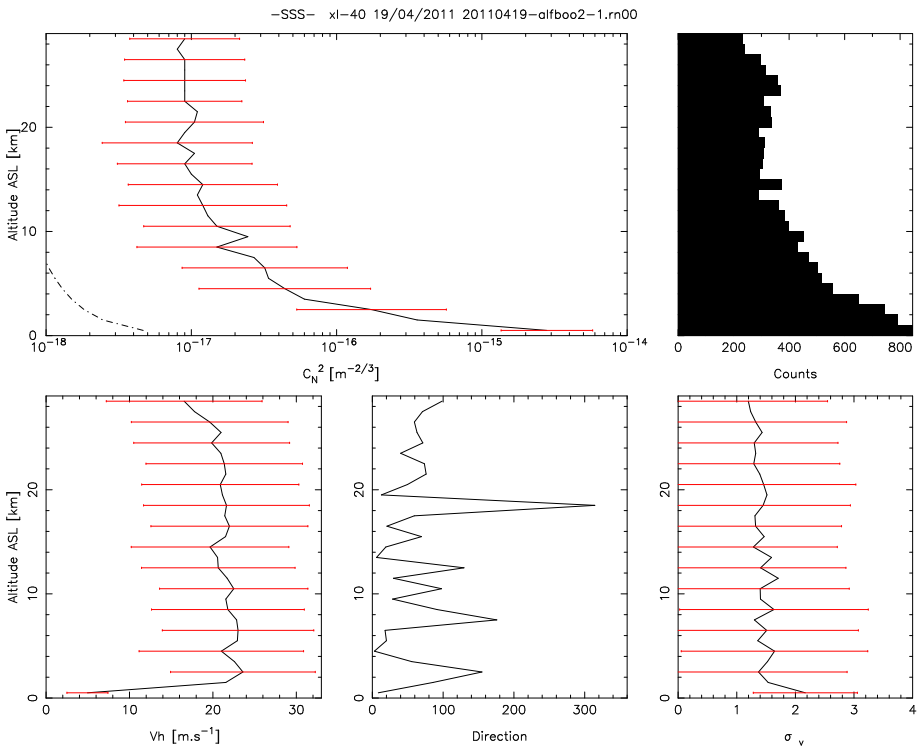}
\end{center}
\caption{\label{label} The profiles of optical turbulence and wind
speed. All profiles were measured by SSS on 19 April 2011, at
Xinglong station.}
\end{figure}

From top left to bottom right, Fig.6 shows the temporal evolution of
the seeing, the coherence volume\cite{Lloyd04}, the isoplanatic
angle, and the time coherence; these key parameters of optical
turbulence can be deduced from the knowledge of optical turbulence
and wind speed profiles.

\begin{figure}
\begin{center}
\includegraphics[width=23pc,angle=-90]{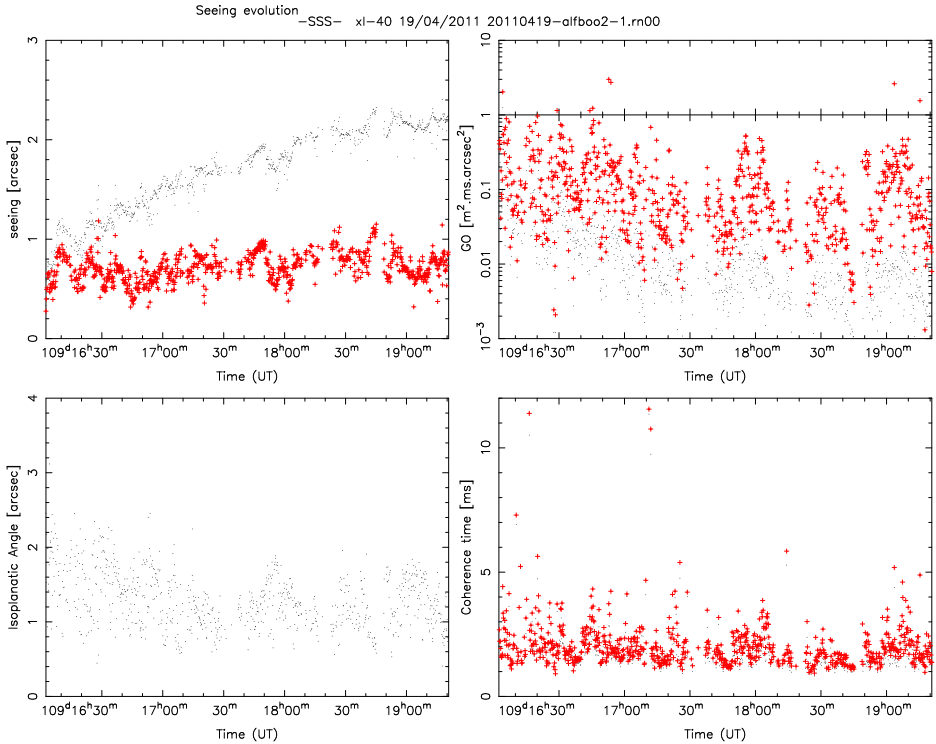}
\end{center}
\caption{\label{label} The temporal evolution of the key parameters
of optical turbulence on 19 April 2011.}
\end{figure}

\section{Conclusion}
We have developed a Single Star Scidar system for astronomical site testing.
The SSS can provide vertical profiles of turbulence intensity and
wind speed from ground up to an altitude of 30$km$.
The SSS can employ only a 40$cm$ telescope, making it portable and equipped
easily at field candidate sites.

There is a plan to upgrade the SSS system and make an auto-SSS system with more friendly
control, so that our SSS can be usable under the formidable conditions at Tibetan high plateau and Dome A, Antarctica.

\section*{Acknowledgements}
This work was supported by the National Natural Science Foundation
of China (Grant Nos. 10903014, 11103042 and 11073031), and by the
Young Researcher Grant of National Astronomical Observatories,
Chinese Academy of Sciences.

\end{document}